\documentclass[12pt]{article}
\usepackage{epsfig,amsfonts,amssymb}
\usepackage{hyperref}
\usepackage{cite}
\input epsf.sty
\usepackage{cite}

\def\ZZZ{{\hbox{ Z\kern-1.6mm Z}}}
\def\RRR{{\hbox{ R\kern-2.4mm R}}}
\def\CCC{{\hbox{ C\kern-2.0mm C}}}
\def\zzz{{\hbox{z\kern-1mm z}}}

\newcommand{\vt}{\vartheta}

\newcommand{\qeq}{{\hbox{=\kern-2.3mm ? \kern.5mm }}}
\renewcommand{\qeq}{=}

\newcommand{\wt}{\widetilde}
\newcommand{\wh}{\widehat}

\newcommand{\NN}{{\cal N}}

\newcommand{\cps}{\Psi}
\newcommand{\crh}{\check\rho}
\newcommand{\cs}{\check\sigma}
\newcommand{\cv}{\check v}

\newcommand{\be}{\begin{equation}}
\newcommand{\ee}{\end{equation}}
\newcommand{\ben}{\begin{eqnarray}\displaystyle}
\newcommand{\een}{\end{eqnarray}}

\newcommand{\bea}[1]{\begin{eqnarray}\label{#1} }
\newcommand{\eea}{\end{eqnarray}}

\newcommand{\refb}[1]{(\ref{#1})}
\newcommand{\p}{\partial}
\newcommand{\sectiono}[1]{\section{#1}\setcounter{equation}{0}}

\def\one{{\hbox{ 1\kern-.8mm l}}}
\def\zero{{\hbox{ 0\kern-1.5mm 0}}}

\begin{document}

\baselineskip 24pt

\begin{center}
{\Large \bf
$\NN=8$ Dyon Partition Function and Walls of Marginal Stability}

\end{center}

\vskip .6cm
\medskip

\vspace*{4.0ex}

\baselineskip=18pt

\centerline{\large \rm   Ashoke Sen }

\vspace*{4.0ex}

\centerline{\large \it Harish-Chandra Research Institute}

\centerline{\large \it  Chhatnag Road, Jhusi,
Allahabad 211019, INDIA}

\vspace*{1.0ex}
\centerline{E-mail:  sen@mri.ernet.in, ashokesen1999@gmail.com}

\vspace*{5.0ex}

\centerline{\bf Abstract} \bigskip

We construct the partition function of 1/8 BPS dyons in type II string
theory on $T^6$ from counting of microstates of a D1-D5 system in
Taub-NUT space. Our analysis extends
the earlier ones by Shih, 
Strominger and Yin and by Pioline by taking into account the 
walls of marginal stability on which
a 1/8 BPS dyon can decay into a pair of half-BPS dyons.
Across these walls the dyon spectrum 
changes discontinuously, and as a result
the spectrum is
not manifestly invariant under S-duality transformation of the 
charges.  However
the
partition
function is manifestly S-duality invariant and takes the same form
in all domains of the moduli
space separated by walls of marginal stability, -- the spectra
in different domains being obtained by choosing different integration
contours along which we carry out the Fourier transform of the
partition function. The jump in the spectrum across a wall of
marginal stability, calculated from the behaviour of the partition function
at an appropriate pole, reproduces the expected wall crossing formula.

\vfill \eject

\baselineskip=18pt

\tableofcontents

\sectiono{Introduction, conventions and summary} \label{sintro}

Since the original proposal of
Dijkgraaf, Verlinde and Verlinde\cite{9607026}
for quarter BPS dyon spectrum
in heterotic string theory compactified on $T^6$, 
there has been
extensive study of dyon spectrum in a variety of 
$\NN=4$ supersymmetric string 
theories\cite{0412287,0505094,
0506249,0508174,0510147,0602254,
0603066,0605210,0607155,0609109,0612011,0702141,
0702150,0705.1433,0705.3874,0706.2363,
0708.1270,0802.0544,0802.1556}. 
Typically in these theories
the
dyon spectrum  jumps discontinuously across walls of marginal
stability in the moduli space on which a quarter BPS dyon can decay
into a pair of half-BPS dyons.
One of the important results
which has emerged out of the recent studies is that 
the dyon partition function does not
change across these walls of marginal stability. 
Instead in order to extract the degeneracy
 -- or more appropriately an index --
 in different
domains bounded by walls of marginal stability,
the partition function needs to be integrated along different
contours. As a result the jump in
the index across
a wall of marginal stability may be determined by evaluating the
residue of the integrand at the pole that the contour crosses as we
deform it from its initial position to the final 
position\cite{0702141,0702150,0705.3874,
0706.2363,0802.0544,0802.1556}. 

Given this result in $\NN=4$ supersymmetric string theories it
is natural to explore if this generalizes to other string theories with
different amounts of supersymmetry. In this paper we shall extend
the analysis to the 1/8 BPS dyon
partition function of $\NN=8$ supersymmetric string
theory obtained by compactifying type IIB string theory on a six
dimensional torus. The dyon spectrum in this theory has been
analyzed before in \cite{0506151,0506228,0508174} 
using 4D-5D lift\cite{0503217} and earlier results
of \cite{9903163} for the partition function of 
D1-D5 system in type IIB string theory
compactified on $T^5$. In our analysis we
include a larger set of dyons charges -- including
those corresponding to configurations with
no D1-branes -- in the definition of the partition
function. 
This new partition function turns out to have properties similar
to those of $\NN=4$ supersymmetric string theories, \i.e.\ the same
partition function, integrated along different contours, gives us
the index of 1/8 BPS dyons in different domains in the moduli space
separated by walls of marginal stability. 
Furthermore, the partition function has manifest S-duality invariance.
We also verify explicitly that the jump in the index across a
wall of marginal stability, computed from the behaviour of the
partition function near an appropriate pole, reproduces the expected
wall crosing formula\cite{0005049,0010222,0101135,0206072,
0304094,0702146,0706.3193,kont,private}.

We shall now give a brief summary of our analysis and the
results. The analysis proceeds exactly
parallel to that for $\NN=4$ supersymmetric string theories
described in \cite{0605210,0607155,0609109,0708.1270}.
We consider type IIB string theory compactified on a six
dimensional torus denoted as $T^4\times S^1\times \wt S^1$ 
and consider in this theory a 1/8 BPS configuration consisting
of a Kaluza-Klein (KK) monopole -- or equivalently
a Taub-NUT space --  associated with the $\wt S^1$
circle, a D5-brane wrapped on $T^4\times S^1$ and $Q_1$ 
D1-branes
wrapped along $S^1$, carrying $-n$ units of momentum along
$S^1$ and $J$ units of momentum along $\wt S^1$. We shall
call this the first description. There is a second description, obtained
from the above configuration 
by applying an $SL(2,\ZZZ)$ duality of the ten dimensional
type IIB string theory, 
followed by an $R\to 1/R$ duality on the circle $\wt S^1$ that
takes it to a dual circle $\wh S^1$ and finally making
a six dimensional string-string duality transformation that
dualizes the NS-NS 2-form field. Using the sign
conventions reviewed in \cite{0708.1270} we find that this maps the
original 
configuration  to 
a KK monopole associated
with the circle $\wh S^1$, $-Q_1$ NS 5-branes wrapped 
on $T^4\times S^1$,
$-J$ NS 5-branes wrapped on $T^4\times \wh S^1$
and a fundamental type II string wrapped
on $S^1$, carrying $-n$ units of momentum along $S^1$.
In the second description,
 we denote by $Q$ and $P$ the 12 dimensional
electric and magnetic charge
vectors in the convention where 
the elementary
strings carry electric charges and the solitons carry
magnetic charges. We also denote
by $L$ the $SO(6,6)$ continuous T-duality
invariant metric of signature (6,6) and define
\be \label{ek2}
Q^2=Q^T L Q, \qquad P^2 = P^T L P, \qquad Q\cdot P
=Q^T L P\, .
\ee
Then for the configuration described above we have\cite{0708.1270}
\be \label{ek3}
Q^2 = 2n, \qquad P^2 = 2 Q_1, \qquad Q\cdot P=J\, .
\ee
In order to develop our analysis in close analogy to that in $\NN=4$
supersymmetric string theories we shall restrict our attention to
charge vectors of the type described above, or those which can be
obtained from them by T-duality and electric-magnetic S-duality transformation
in the second description of the theory. 
This allows us to have charges associated with momentum and
fundamental string winding numbers along the six circles, and also
Kaluza-Klein monopole and H-monopole charges associated with
these circles, but does not allow us to have D-brane charges.
We shall also restrict the moduli of the theory to lie in a subspace
where no RR sector fields in the second description 
are switched on. 
This corresponds to requiring the states and the moduli to be
invariant under $(-1)^{F_L}$ (or $(-1)^{F_R}$), and is a consistent
set of restrictions. In this restricted
subspace the duality group is a product of the
discrete T-duality group 
$SO(6,6;\ZZZ)$ and the discrete S-duality group $SL(2,\ZZZ)$.
Since we shall be computing an index 
which does not change continuously
under variation of the moduli, we expect
our results to be valid at least within an open neighbourhood of this
restricted subspace in the full moduli space.\footnote{Due
to the restriction on the charges our formula for the index is not
manifestly invariant under the U-duality group
$E_{7(7)}(\ZZZ)$. However we expect that it can be regarded as
the special case of a more general formula which is manifestly
U-duality invariant. 
As we shall see later, for $Q^2 P^2 - (Q\cdot P)^2<0$ there is
an additional complication due to the existence of walls of
marginal stability, -- in this case the index not only depends on the
charges but also on the domain of the moduli space where we compute
the index.}

A 1/8 BPS state of the theory
carrying the charges $(Q,P)$ given above
breaks 28 of the 32 supersymmetry generators of the theory. As a 
result it is accompanied by 28 real fermion zero modes and quantization
of these fermion zero modes gives a $2^{14}$-fold degenerate state.
A generic 1/8 BPS state is obtained by taking a tensor product
of this basic supermultiplet with a supersymmetry singlet state. We shall
call such a state 
bosonic or fermionic depending on whether this
supersymmetry singlet state is bosonic or fermionic, and denote by
$d(Q,P)$ the number of bosonic supermultiplets minus the number of
fermionic supermultiplets carrying charges $(Q,P)$. This number can
be calculated from the helicity supertrace\cite{9708062,9708130}
\be \label{ek4}
-B_{14} =  -{1\over 14!}\,
Tr_{Q,P}(-1)^{F} (2h)^{14}\, ,
\ee
where $h$ denotes the helicity of a state, $F$ denotes the
fermion number and $Tr_{Q,P}$ denotes 
trace over all states of charge $(Q,P)$. The need for the 
$(2h)^{14}$ factor may be understood as follows.
For every pair of fermion
zero modes we have a 2-fold degeneracy with the two states
carrying opposite fermion number and differing in helicity by $1/2$.
Thus $Tr(-1)^{F}$ vanishes for this pair
of states and we need to insert
a factor of $2h$ into the trace to get a non-zero answer $i$. 
Since there are altogether 28 fermion zero modes, we need to
insert 14 factors of $2h$ to `soak up' all the fermion zero modes
and give a non-vanishing answer $(i)^{14}=-1$.
This
explains the need for the $(2h)^{14}$ factor. Since all the factors of
$2h$ are soaked up by the fermion zero modes associated with
the broken supersymmetry generators, \refb{ek4} effectively evaluates
$Tr(-1)^F$ on the supersymmetry singlet states with which we tensor
the
basic 1/8 BPS supermultiplet. The $1/14!$ factor accounts for the
fact that the 14 factors of $2h$ may be distributed among the 
traces over the 14 pairs of fermion zero modes in $14!$ different ways.

For the dyon configurations described earlier,
$d(Q,P)$ may be regarded as a function 
$f(n,Q_1,J)$
of $n=Q^2/2$,
$Q_1=P^2/2$ and $J=Q\cdot P$. 
It now follows from the analysis of \cite{0712.0043, 0801.0149}
that the index $-B_{14}$ of any other 1/8 BPS dyon
in the restricted subspace, carrying charges $(Q,P)$, 
is given by $f(Q^2/2,P^2/2,Q\cdot P)$ as long as 
$\gcd(Q\wedge P)=1$.
We define the partition function
for these dyons to be
\ben \label{ek4.1}
\cps(\crh,\cs,\cv) 
&=& \sum_{n,Q_1,J} (-1)^{J+1}\,
f(n,Q_1,J) \, e^{2\pi i (n\cs + Q_1\crh
+J\cv)} \nonumber \\
&=&\sum_{Q^2, P^2,Q\cdot P}
(-1)^{Q\cdot P+1} \, f(Q^2/2,P^2/2,Q\cdot P)\, 
\, e^{i\pi (\cs Q^2 + \crh P^2 
+ 2 \cv Q\cdot P)}\, .
\een
Conversely, we may write
\be \label{ek5}
d(Q, P)
={(-1)^{Q\cdot P+1}}
\int_{i M_1-1/2}^{iM_1+1/2} d\crh
\int_{iM_2-1/2}^{i M_2+1/2} d\cs 
\int_{i M_3-1/2}^{i M_3+1/2} d\cv \, 
e^{-i\pi (\cs Q^2 + \crh P^2 
+ 2 \cv Q\cdot P)} \, 
\cps(\crh, \cs, \cv)\, ,
\ee
where $M_1$, $M_2$ and $M_3$, describing the imaginary
parts of $\crh$, $\cs$ and $\cv$, need to be fixed to values where
the sum in \refb{ek4.1} is convergent.  

There is one subtle issue that needs to be mentioned here.
The full $\NN=8$ supersymmetry transformations acting on the
fields in the restricted subspace of the second description,
where we set the RR sector moduli and the D-brane charges to zero,
will take us out of this subspace.
However there are two different $\NN=4$ subalgebras which
preserve this subspace, -- one acting on the right-moving world-sheet
fields and the other acting on the left-moving world-sheet fields.
We shall call these right-handed and the left-handed supersymmetry
algebras, --
these commute with
$(-1)^{F_L}$ and $(-1)^{F_R}$ symmetries respectively.
A 1/8 BPS state of the full theory without any RR charge, 
having four unbroken 
supersymmetries, must correspond to a quarter BPS state of one
of these two $\NN=4$ supersymmetry algebras. For definiteness
we shall focus on those states which preserve quarter of the
supersymmetries of the right-handed supersymmetry algebra.
In particular in defining the partition function we shall sum over only
states of this type, and not the ones which preserve quarter of the
supersymmetries of the left-handed supersymmetry algebra. 
This corresponds to summing over states with $Q_1,n\ge 0$.
We shall see however that the same partition function, expanded
in a different way, also captures information about quarter BPS states
of the left-handed supersymmetry algebra.

We compute the partition function $\cps$ by counting microstates
carrying charge quantum numbers $(Q,P)$ in the weakly coupled
type II string theory
in the first description.
Our result for $\cps$ is
\be \label{ek6}
\cps(\crh,\cs,\cv) = \sum_{k\ge 0} \, \sum_{l\ge 0}\,
\sum_{j\in \zzz\atop j>0\, \hbox{{\small for}}\, k=l=0}
\left( 1- e^{2\pi i (l\cs + k\crh
+j\cv)}\right)^{-2} \, e^{2\pi i (l\cs + k\crh
+j\cv)}\, \wh c(4k l - j^2)\, ,
\ee
where $\wh c(u)$ is defined through the 
relations\cite{9903163,0506151}
\be \label{ek6.5}
-\vt_1(z|\tau)^2 \, \eta(\tau)^{-6} \equiv \sum_{k,l} \wh c(4k-l^2)\, 
e^{2\pi i (k\tau+l z)}\, .
\ee
$\vt_1(z|\tau)$ and $\eta(\tau)$ are respectively the odd Jacobi
theta function and the Dedekind eta function. 
The $k\ge 1$ terms in the sum in \refb{ek6}
are identical to the partition function
of the D1-D5 system in type IIB string theory 
compactified on $T^4\times
S^1$\cite{9903163}. 
As we shall explain, the $k=0$ term comes from the dynamics of
a single D5-brane moving in the KK monopole background
and is essential for S-duality invariance of the partition function and
consistency with the wall crossing formula.
In this context we also note that in the $k=l=0$ term if we had
put the restriction $j<0$ instead of $j>0$ we would have gotten the
same analytic function.

The partition function \refb{ek6}  itself does not 
give the index $d(Q,P)$
unambiguously; we must specify the values of $M_1$, $M_2$
and $M_3$ appearing in \refb{ek5} or equivalently describe how we
should Fourier expand $\cps$ to extract the index. It turns out that
in the weakly coupled type IIB string theory in the first description
where we have carried out the computation,
we need
to first expand $\cps$ in powers of $e^{2\pi i\crh}$ and $e^{2\pi i\cs}$
and then expand the coefficient of each term in powers of
$e^{2\pi i \cv}$ or $e^{-2\pi i\cv}$. This unambiguously determines
the contribution from the $(k,l)\ne (0,0)$ terms, -- for any given
power of $e^{2\pi i\crh}$ and $e^{2\pi i\cs}$ the power of
$e^{2\pi i\cv}$ is bounded both from above and below and the Fourier
expansion in $e^{2\pi i\cv}$ is unambiguous. However for the
$k=l=0$ term the summand can be expanded either as a power series
expansion in $e^{2\pi i \cv}$ or as a power series expansion in
$e^{-2\pi i\cv}$, with infinite number of terms in each case.
The correct choice depends on whether the
angle $\theta$
between $S^1$ and $\wt S^1$ is larger than $90^\circ$ or less
than $90^\circ$. 
In other words, the actual value of the index changes as we cross
the $\theta=90^\circ$ line.
This is entirely analogous to the corresponding
result found in \cite{0605210} and reviewed in detail in
\cite{0708.1270} for $\NN=4$ supersymmeric string theories.
Equivalently, while determining $d(Q,P)$ via eq.\refb{ek5},
the quantities $M_1$, $M_2$ and $M_3$ which fix the integration
contour need to be chosen in the range
\be \label{ek8}
M_1,M_2>> |M_3|>>0\, ,
\ee
with the sign of $M_3$ determined by the angle between $S^1$ and
$\wt S^1$. If we try to deform the contour for $M_3>0$ to the one for
$M_3<0$ we pick up the residue
at the pole of $\cps(\crh,\cs,\cv)$ at $\cv=0$,
and as a result the index changes by an amount determined from
the residue at the pole. This correctly accounts for the 
change in the index across the $\theta=90^\circ$ line.

Our analysis determines the form of the partition function in a
specific domain -- more precisely two domains
separated by the $\theta=90^\circ$ line --
of the moduli space corresponding to weak string
coupling in the first description of the theory.
{\it A priori} the index $d(Q,P)$ and the partition function $\cps$
could have different forms in other domains in the moduli space
separated by walls of marginal stability. However 
as in the case of all known
examples in $\NN=4$ supersymmetric string theories,
we find that the 
partition function of the $N=8$ supersymmetric string theory
is also described by the same
analytic function in all domains of the moduli space. The
choice of the integration contour, encoded in the choice of $M_1$,
$M_2$ and $M_3$, differs in different domains separated
by the walls of marginal stability. 
The proof of this follows the same logic as in the case of
$\NN=4$ supersymmetric string theories\cite{0702141,0708.1270}.
First of all we note that S-duality transformations take us
from one domain bounded by walls of marginal stability
to another such domain\cite{0702141,0702150}. In particular
if the original domain is bounded by walls of marginal stability on
which the dyon of charge $(Q,P)$ decays into dyons of charge
$(\alpha_iQ+\beta_iP, \gamma_iQ+\delta_iP)$ and
$((1-\alpha_i)Q-\beta_iP, -\gamma_iQ+(1-\delta_i)P)$, then S-duality
transformation of the charges and the moduli
by the $SL(2,\ZZZ)$ matrix $\pmatrix{a&b\cr c & d}$
maps it into a new domain in which $\pmatrix{\alpha_i&\beta_i\cr
\gamma_i&\delta_i}$ is replaced by\cite{0702141,0801.0149}
\be\label{ealbe}
\pmatrix{\alpha_i'&\beta_i'\cr
\gamma_i'&\delta_i'}= \pmatrix{a&b\cr c & d}
\pmatrix{\alpha_i&\beta_i\cr
\gamma_i&\delta_i}\pmatrix{a&b\cr c & d}^{-1}\, .
\ee
Thus we can find the index for the
dyons in other domains by an S-duality transformation of our
original formula.  On the other hand explicit S-duality transformation
of \refb{ek5} by the $SL(2,\ZZZ)$ matrix
$\pmatrix{a  & b\cr c & d}$
leads to a new formula for the index where we replace
in \refb{ek5} $\cps(\crh,\cs,\cv)$ and $(M_1,M_2,M_3)$ 
by\cite{0702141,0802.0544}
\be \label{ereplace}
\cps(d^2 \crh + b^2 \cs + 2 bd \cv,
c^2 \crh + a^2 \cs + 2 ac \cv, 
cd \crh + ab \cs + (ad+bc) \cv)
\ee
and
\be \label{ereplace2}
(a^2 M_1 + b^2 M_2 - 2 ab M_3,  
c^2 M_1 + d^2 M_2 - 2 cd M_3, 
- ac M_1 - bd M_2 + (ad + bc) M_3)
\ee
respectively.
Thus if we can show 
that the partition function $\cps$ is manifestly invariant
under the S-duality transformation:
\be \label{esdual}
\cps(\crh,\cs,\cv) = \cps(d^2 \crh + b^2 \cs + 2 bd \cv,
c^2 \crh + a^2 \cs + 2 ac \cv, 
cd \crh + ab \cs + (ad+bc) \cv), 
\quad \pmatrix{a & b\cr c & d}\in SL(2,\ZZZ) \, ,
\ee 
then the result for the index in different domains in the moduli space
will be given just by changing the integration contour according to
\refb{ereplace2}.
We prove \refb{esdual} explicitly in \S\ref{sdual}. 

In order to use \refb{ereplace2} to find the integration contour
in different domains in the moduli space we need information about
the choice of contour in at least one such domain. This can be gotten
from the results of weak coupling calculation in the first description.
In particular if we consider the domain bounded by the walls
corresponding to the decays $(Q,P)\to (Q,0)+(0,P)$, 
$(Q,P)\to (Q,Q)+(0, P-Q)$ and $(Q,P)\to 
(Q-P,0)+(P,P)$, then we need to choose
the $M_i$'s as 
$M_1,M_2>> 
-M_3>>0$\cite{0702141,0708.1270}.\footnote{Although we are
using these decays to label different domains in the
moduli space, we shall see in
\S\ref{swall} that the actual jump in the index across
many
of these walls vanishes in the $\NN=8$ supersymmetric string theory.}
An explicit prescription for $(M_1,M_2,M_3)$ at different points
in the moduli space satisfying \refb{ereplace2} can be found
in \cite{0706.2363,0802.0544}.

The S-duality invariance of
$\cps$ allows us to calculate the jump in the index across
a wall of marginal stability by calculating the residue at the pole(s)
the contour crosses as we deform it from the initial position to the
final position calculated according to \refb{ereplace2}.
It follows from the corresponding analysis
in $\NN=4$ supersymmetric
string theories\cite{0702141,0802.0544,0802.1556} that 
as we cross a wall of marginal stability associated with the
decay into a pair of half-BPS states\footnote{It follows 
from the analysis of
\cite{0702141,0707.1563,0707.3035,0710.4533} that in the
restricted subspace of the moduli space we are considering, such
decays occur on codimension 1 subspaces. 
This is a necessary (but not sufficient) condition for an index to
jump across this subspace, since for higher codimension subspaces
we can avoid the change in the index by 
{\it going around the subspace}.
We shall see later that for a subset of the
decays of the type described in \refb{e2}-\refb{e3} for which
$(\alpha Q+\beta P)$ and $(\delta Q - \beta P)$ are null and as
a result
the final decay products are half-BPS under the
full $\NN=8$ supersymmetry algebra and not just the $\NN=4$
subalgebra that we are considering here, 
the index $B_{14}$ jumps
discontinuously. Thus we expect that such
decays will continue to occur on a codimension one subspace of the
full moduli space even after turning on the RR sector moduli.
For the other decays of the type described in \refb{e2}-\refb{e3}
for which either $(\alpha Q+\beta P)$ or $(\delta Q - \beta P)$
fails to be null, the change in $B_{14}$
vanishes. Thus the associated
subspace may or may not
remain codimension one in the full moduli space. Finally,
if we consider decays which fail to satisfy \refb{e2.55} and/or
\refb{e3},  then
it follows from the analysis of \cite{0702141,0707.1563,0707.3035,0710.4533} that such decays occur
on subspaces of codimension $>1$ in the restricted subspace of the
moduli space we are considering and hence also in the full
moduli space. The change in $B_{14}$ associated with such decays
always vanishes.}
\be \label{e2}
(Q,P) \to (Q_1,P_1)+(Q_2,P_2)\, ,
\ee
\be \label{e2.55}
(Q_1,P_1) = (\alpha Q +\beta P, \gamma Q +\delta P), 
\qquad (Q_2,P_2)=
(\delta Q -\beta P, -\gamma Q + \alpha P)\, ,
\ee
\be \label{e3}
\alpha\delta=\beta\gamma, \qquad \alpha+\delta=1, \qquad
\alpha,\beta,\gamma,\delta\in\ZZZ\, ,
\ee
the contour crosses the pole at
\be \label{e4}
\crh \gamma - \cs \beta + \cv (\alpha-\delta) = 0\, .
\ee
Thus the jump in the index across a wall of marginal
stability is given by the residue of the
integrand in \refb{ek5} at the pole at \refb{e4}.
We  check  in
\S\ref{swall} that the jump in the index predicted by this
formula at the wall associated with the decay $(Q,P)\to
(Q,0)+(0,P)$
reproduces the expected change
in the index across a wall of marginal 
stability\cite{0005049,0010222,0101135,0206072,0304094,0702146,
0706.3193,kont,private}. Similar agreement at the other walls of
marginal stability then follows from S-duality invariance
of $\cps$.

Refs.\cite{0506151,0506228} proposed formul\ae\ similar to the one
given in \refb{ek6} for the 1/8 BPS dyon spectrum in type IIB string
theory on $T^6$. However both these proposals treated the dyon
spectrum as universal independent of the region of the moduli space
where we calculate the spectrum. Thus in any given region in the
moduli space these proposals differ from ours in a subtle way. 
As we shall see below eq.\refb{ejump}, the index jumps
discontinuously across a wall of marginal stability only for states
with $Q^2P^2-(Q\cdot P)^2<0$. Thus the subtle difference between
our results and those proposed in \cite{0506151,0506228}
arises only for such states.

Since
for positive
$Q^2P^2-(Q\cdot P)^2$ we do not
have any associated wall of marginal stability, the
index calculated from \refb{ek5}, \refb{ek6} is independent of
the choice of $(M_1,M_2,M_3)$ and 
the
formula for the index $d(Q,P)$ takes the 
form\cite{9903163,0506228}
\be \label{ei1}
d(Q,P)=(-1)^{Q\cdot P+1}
\sum_{s|Q^2/2,P^2/2,Q\cdot P; s>0} \, s\, \wh c\left(
{Q^2 P^2 - (Q\cdot P)^2\over s^2}\right)\, .
\ee
This formula is manifestly S-duality invariant since
$Q^2 P^2 - (Q\cdot P)^2$,
$\gcd(Q^2/2,P^2/2,Q\cdot P)$ and $(-1)^{Q\cdot P}$
are S-duality invariants.

The partition function \refb{ek6} has been derived for the 1/8 BPS
states whose unbroken supersymmetries lie in the right-handed
$\NN=4$ supersymmetry algebra. The index for 1/8 BPS states whose
unbroken supersymmetries lie in the left-handed $\NN=4$
supersymmetry algebra can be obtained from the former by world-sheet
parity transformation exchanging these two supersymmetry algebras.
This effectively replaces the $O(6,6)$ invariant metric
$L$ by $-L$ in all formul\ae.
In particular this 
changes 
$(Q^2,P^2,Q\cdot P)$ to $(-Q^2,-P^2,-Q\cdot P)$.
Thus the new partition function is related to the previous one
by the transformation $(\crh,\cs,\cv)\to (-\crh,-\cs,-\cv)$.
This however produces the same analytic function $\cps(\crh,\cs,\cv)$
since each term in the sum in \refb{ek6} is invariant under the
transformation $(\crh,\cs,\cv)\to (-\crh,-\cs,-\cv)$. 
Thus the same partition function $\cps(\crh,\cs,\cv)$
contains information about both types of 1/8 BPS states, but the
choice of $(M_1,M_2,M_3)$ in \refb{ek5} for these two types
of dyons are related by the transformation $(M_1,M_2,M_3)\to
(-M_1,-M_2,-M_3)$, accompnied by $L\to -L$ replacement
in the expression for $M_i$'s in terms of the
moduli given in \cite{0706.2363,0802.0544}. 
Also for a given charge vector $(Q,P)$,
the physical location
of the walls of marginal stability in the moduli space, 
computed in
\cite{0702141}, will  differ for the two
algebras due to the $L\to -L$ replacement in the associated
formul\ae.

Finally we note that some of the $\NN=2$ supersymmetric models
discussed in \cite{0711.1971} have properties close to that of
$\NN=4$ supersymmetric string theories. It will be
interesting to explore if the phenomenon of having a universal
partition function independent of the domains in the moduli space
holds for these theories as well.

\sectiono{Microstate counting} \label{smicro}

In this section we shall describe how we arrive at the formula
\refb{ek6} for the dyon partition function. We shall carry out our
analysis in the weak coupling limit of the first description of the
theory, regarding it as a collection of KK monopole
--  D1-D5 system
carrying momenta along the circles
$S^1$ and $\wt S^1$. As in \cite{0605210}, 
in this limit we can regard
the degrees of freedom of the
KK monopole, the D1-D5 center of mass motion
and the D1-D5 relative motion as three non-interacting systems
and calculate the partition function of the combined system by taking
the product of the three partition functions.

Since we are computing the partition function of 1/8 BPS states,
special care must be taken to ensure that all the fermion zero
modes associated with the 28 broken supersymmetry generators
are soaked up. As described in \S\ref{sintro}, each pair of fermion
zero modes are soaked up by an insertion of $(-1)^F 2h$ factor in the
helicity trace. Let us examine how the fermion zero modes get
distributed between the three subsystems described above. First of
all the KK monopole associated with
$\wt S^1$ breaks 16 of the 32 supersymmetries of the
original theory.
Thus the world-volume theory on the KK monopole
has 16 fermion zero modes which must be soaked up by a factor
of $(-1)^F (2h)^{8}$ inserted into the partition function. 
The classical D5-brane in the KK monopole background breaks
8 of the remaining 16 supersymmetries and hence its
world-volume carries 8 fermion zero modes. We need a
factor of $(-1)^F (2h)^4$ inserted into the partition function 
of the D5-brane to
soak up these zero modes. Finally it is known from the analysis
of \cite{9903163} that the system 
describing the dynamics of the D1-branes
inside the D5-brane has four fermion zero modes and hence requires
a factor of $(-1)^F (2h)^2$ inserted into the partition function
to soak up these zero modes.
This determines how the 14 factors of $(2h)$ are distributed among
different components of the partition function.

The above analysis also shows that in order to get a non-vanishing
result for the index we must keep both the KK monopole
and the D5-brane in their ground state. In particular we cannot
excite any mode carrying momentum along $S^1$. If we did, then
such a system will break more supersymmetries and we shall need
additional factors of $(2h)$ to be inserted into the partition function
in order to get a non-vanishing result. Since the total power of $2h$ is
fixed to be 14, this would require 
removing some factors of $2h$ from
some other component of the partition function, 
making the corresponding
contribution vanish.\footnote{This should be contrasted with the
corresponding analysis in $\NN=4$ supersymmetric string theories
where  the KK monopole and the D5-brane
can have excitations carrying left-moving momentum along
$S^1$ without breaking any
additional supersymmetry.}
Thus the non-trivial contribution to the 
partition function comes solely from the relative motion of the D1-D5
system. This was evaluated in \cite{9903163}, 
yielding the answer\footnote{The $(-1)^{Q\cdot P}$ 
factor in eqs.\refb{ek4.1},
\refb{ek5} relating the index to the partition function
has the same origin as in the case of $\NN=4$
supersymmetric string 
theories\cite{0508174,0706.2363,0708.1270} and will not be discussed
here.}
\be \label{el1}
\sum_{k\ge 1} \, \sum_{l\ge 0}\,
\sum_{j\in \zzz}
\left( 1- e^{2\pi i (l\cs + k\crh
+j\cv)}\right)^{-2} \, e^{2\pi i (l\cs + k\crh
+j\cv)}\, \wh c(4k l - j^2)\, ,
\ee
where $\wh c(u)$ has been defined in eq.\refb{ek6.5}. This can be
identified with the contribution to \refb{ek6} from the
$k\ge 1$ terms.

In arriving at \refb{el1} we have implicitly assumed that 
leaving aside the
degeneracy implied by the 8 broken supersymmetry generators,
there is a 
unique supersymmetric state describing the D5-brane bound to the
KK monopole.
Is this
true? A D5-brane in flat space has altogether 16 fermion fields on its
world-volume. As pointed out before,
classically a D5-brane in the background of the
KK monopole has eight unbroken supersymmetries and breaks eight
supersymmetries. 
Thus the world-volume theory of the D5-brane
should contain eight free goldstino fermions. These, together with the
four scalar fields associated with the Wilson lines along $T^4$ describe
a free (4,4) supersymmetric field theory in 1+1 dimensions spanning the
time coordinate and the coordinate along $S^1$. 
The rest of the
eight fermion fields on the D5-brane world-volume combine with the
four scalar fields describing motion along the Taub-NUT space
to give an interacting (4,4)
supersymmetric sigma model with Taub-NUT target
space. 
By the standard argument the
number of supersymmetric ground states of this system is equal to
the number of harmonic forms on the Taub-NUT space which is
1\cite{brill,pope}. 
Since the harmonic form is normalizable the state describes
a bound state\cite{9601085,9601097}. 
This confirms our implicit assumption that 
leaving aside the
degeneracy implied by the 8 broken supersymmetry generators,
there is a 
unique supersymmetric state describing the D5-brane bound to the
KK monopole.

The above analysis holds when the $S^1$ and $\wt S^1$
circles are orthogonal. When we switch on a modulus that
changes the angle between $S^1$ and $\wt S^1$ there is an
attractive force between the D5-brane and the KK monopole,
giving rise to additional potential terms\cite{0605210}. 
Since in the limit in which
the size of the Taub-NUT space is large this potential term 
as well as the Taub-NUT metric affects mainly
the modes  independent of the $S^1$ coordinate,
we can treat the non-zero mode oscillators carrying
momentum along $S^1$ as free and
focus our attention on the supersymmetric
quantum mechanics describing the zero modes.
The latter is obtained via
dimensional reduction of the 1+1 dimensional field theory
to 0+1 dimensions.
This system was analyzed in \cite{9907090,0005275,0609055} 
in the context of
quarter BPS dyon spectrum in the $N=4$ supersymmetric gauge
theories and the result may be summarized as follows.
Besides the fully supersymmetric ground state described above
the system has a set of states where 4 of the eight remaining
supersymmetries are broken. Depending on the angle
between $S^1$ and $\wt S^1$ these partially supersymmetric
states have either only positive or only negative
momentum along $\wt S^1$, and there are precisely $|j|$ states
carrying $\wt S^1$ momentum $j$.\footnote{We are counting
only those states which, from the space-time viewpoint, 
preserve quarter of the supersymmetries of the
right-handed $\NN=4$ supersymmetry algebra.
There are also states preserving quarter of the supersymmetries of
the left-handed supersymmetry algebra; for these states the sign
of $\wt S^1$ momentum is opposite.}
Thus the partition
function for these dyons can be written as
\be \label{el2}
\sum_{j=1}^\infty j e^{\pm 2\pi i \cv j} =
 e^{2\pi i v} / (1-e^{2\pi i v})^2\, .
 \ee
 Note that the result for the partition function is independent of
 whether we use positive or negative momentum along $\wt S^1$.
 
 The partially supersymmetric states described above
 are not relevant for the computation at hand
 since due to the four additional broken supersymmetries we need to
 insert a factor of $(-1)^F (2h)^2$ into the partition function. This
 effectively uses up all the factors of $(2h)$ in order to saturate the
 fermion zero modes of the KK monopole - D5-brane system, and
 there is no left over factors of $2h$ which we can insert into the
 partition function of the D1-D5 system. 
 As a result the contribution from
 such terms vanish.

We must note however that in computing the full partition function
we must also include states with $Q_1=0$, \i.e.\ no D1-branes if
there are 1/8 BPS states in this sector.
Proceeding as before we can see that the KK monopole soaks
up a factor of $(2h)^8$ and the 8 broken supersymmetries associated
with the classical 
D5-brane soak up another factor of $(2h)^4$. Thus we are
left with a factor of $(2h)^2$. However unlike in the $Q_1\ne 0$ case
where these factors needed to be inserted in the D1-D5 partition
function to get a non-vanishing contribution,
here we are free to use them to either get additional
excitations on the D5-brane from the left-moving
modes\footnote{Exciting the right-moving modes will break the
right-handed $\NN=4$ supersymmetry algebra but preserve the
left-handed $\NN=4$ supersymmetry algebra.
As mentioned in
\S\ref{sintro}, we shall not include such states in the definition of the
partition function even though they describe 1/8 BPS dyons.} 
of the D5-brane world-volume
theory carrying negative  momenta
along $S^1$,
keeping the zero-mode part of the theory
in the fully supersymmetric ground
state, or keeping the non-zero modes in their ground state
and using one of the partially supersymmetric 
bound states of the zero-mode part of the D5-brane world-volume
theory. The contribution from the first set of states can be
computed by recalling from 
\cite{9903163} that the
left-moving fields on the D5-brane world-volume
have the following 
$(j,2h)$ assignment: four scalars with
$(j,2h)=(0,0)$ representing the Wilson lines along
$T^4$, one scalar each  for
$(j,2h)=(1,1)$, $(1,-1)$, $(-1,1)$ and $(-1,-1)$ representing
the coordinates along the Taub-NUT space,  and 2
fermions each for $(j,2h)=(1,0)$, $(-1,0)$, $(0,1)$ and $(0,-1)$.
Following \cite{9903163} this
yields the answer
\ben \label{el4}
&&  \hskip -20pt
\left.
{1\over 2}
{d^2\over d\wt y^2} \prod_{l=1}^\infty \left[(1-q^l)^{-4}
\left\{
\prod_{j=\pm 1}
 \prod_{\wt j =\pm 1} (1 - q^l y^j \wt y^{\wt j})^{-1}
\right\} \left\{\prod_{j=\pm 1} (1-q^l y^j)^2 
\right\} \left\{\prod_{\wt j =\pm 1}
(1-q^l\wt y^{\wt j})^2
\right\}
\right]
\right|_{\wt y=1}, \nonumber \\
&& 
\qquad \qquad
q \equiv e^{2\pi i\cs}, \quad y\equiv e^{2\pi i\cv}\, .
\een
Explicit evaluation of this term leads to
\be \label{el5}
\sum_{l=1}^\infty \left[ - 2 \left(1 - e^{2\pi i l\cs }\right)^{-2} 
\, e^{2\pi i l\cs }
+ \sum_{j=\pm 1} 
\left(1 - e^{2\pi i (l\cs + j\cv)}\right)^{-2} 
\, e^{2\pi i (l\cs + j\cv)}\right]\, .
\ee
Using the fact that $\wh c(u)=0$ for $u<-1$ and $\wh c(0)=-2$,
$\wh c(-1)=1$,
\refb{el5} can be identified with the $k=0$, $l\ge 1$ terms in
the sum in \refb{ek6}.

Finally we turn to the contribution to the
partition function from the partially supersymmetric bound states of
the D5-brane to the KK monopole.
To evaluate this contribution we note that
since the extra factor of $(2h)^2$ is now absorbed by the
supersymmetric quantum mechanics describing the D5-brane
motion in the KK monopole background, we 
need to evaluate the trace over the non-zero mode oscillators
without any factor of $2h$. The corresponding contribution is
given by \refb{el4} without 
the ${1\over 2}\p_{\wt y}^2$ operation.
At $\wt y=1$ this reduces to 1, indicating that we
cannot excite
any of the oscillators carrying momentum along $S^1$.
Thus the contribution to the partition function from this set
of states is given by \refb{el2}.
This can be identified as the
$k=0$, $l=0$ term in \refb{ek6}.

The sum of the expressions given in \refb{el1}, 
\refb{el5} and \refb{el2}
gives the complete contribution to the partition function stated in
\refb{ek6}.

\sectiono{S-duality invariance of the partition function} \label{sdual}

In this section we shall give a proof of S-duality invariance of the
partition function encoded in \refb{esdual}. This requirement may
be expressed as 
\be \label{esd1}
\cps(\crh,\cs,\cv) = \cps(\crh',\cs',\cv')\, ,
\ee
where
\be \label{esd2}
\crh'=d^2 \crh + b^2 \cs + 2 bd \cv, \quad \cs'=
c^2 \crh + a^2 \cs + 2 ac \cv,  \quad \cv'=
cd \crh + ab \cs + (ad+bc) \cv, 
\quad \pmatrix{a & b\cr c & d}\in SL(2,\ZZZ) \, .
\ee 
The proof of \refb{esd1} goes as follows. We first note that
\be \label{esd3}
k\crh+l\cs+j\cv = k'\crh'+l'\cs'+j'\cv'\, ,
\ee
where
\be \label{esd4}
k'=a^2 k + c^2 l - acj, \quad l'=b^2 k +d^2 l -bdj, \quad
j'=-2abk-2cdl + (ad+bc) j\, .
\ee
Furthermore we have
\be \label{esd5}
4k'l'- j^{\prime 2} = 4kl-j^2\, .
\ee
Using these relations in \refb{ek6} we get
\be \label{esd6}
\cps(\crh,\cs,\cv) = 
\sum_{k\ge 0} \, \sum_{l\ge 0}\,
\sum_{j\in \zzz\atop j>0\, \hbox{{\small for}}\, k=l=0}
\left( 1- e^{2\pi i (l'\cs' + k'\crh'
+j'\cv')}\right)^{-2} \, e^{2\pi i (l'\cs' + k'\crh'
+j'\cv')}\, \wh c(4k' l' - j^{\prime 2})\, .
\ee
If we can now show that $k'$, $l'$ and $j'$ take values over
the same range as 
$k$, $l$ and $j$, then the right hand side of \refb{esd6} can
be identified as $\cps(\crh',\cs',\cv')$. This would prove the desired
relation \refb{esd1}.
We shall prove this separately for the S-duality transformations
$S=\pmatrix{0 & 1\cr -1 & 0}$ and $T=\pmatrix{1 & 1\cr 0 & 1}$.
Since these two matrices generate the whole $SL(2,\ZZZ)$ group,
once we have proven invariance of $\cps$ under $S$ and $T$, it proves
invariance of $\cps$ under the full $SL(2,\ZZZ)$ S-duality group.

First consider the transformation by $S$. In this case \refb{esd4}
takes the form
\be \label{esd7}
k'=l, \quad l'=k, \quad j'=-j\, .
\ee
Thus $k'$ and $l'$ run over non-negative integers. Furthermore
for $(k',l')\ne (0,0)$, $j'$ can take arbitrary integer values,
whereas for $k'=l'=0$, $j'$ takes only negative integer values. Using
$j'\to -j'$ symmetry of the summand in \refb{esd6} for $k'=l'=0$
we can turn this into sum over positive integer values of $j'$. Thus
the range of $(k',l',j')$ coincides with the range of summation over
$(k,l,j)$. This proves
invariance of $\cps$ under the transformation $S$.

We now turn to the transformation $T$. In this case \refb{esd4}
gives
\be \label{esd8}
k'=k, \quad l'=k+l-j, \quad j'=j-2k\, .
\ee
Using the $j\to -j$ invariance of the $k=l=0$ term
in \refb{ek6} we shall take
the sum over the original variable $j$ to run over negative integer
values. 
We now see from \refb{esd8} that $k'$ always
takes non-negative
integer values. This agrees with the corresponding range of values
of $k$. To determine the range of values taken by $l'$ we shall
consider two cases separately. First consider the case $k=l=0$.
In this case $l'=-j$. Since $j<0$, we see that $l'$ takes positive
integer values. On the other hand for $(k,l)\ne (0,0)$  we have
\be \label{esd9}
(k+l)^2 - j^2 = (k-l)^2 +4kl-j^2 \ge -1\, ,
\ee
where in the last step we have
used the
fact that $j^2 \le 4 kl+1$ for $\wh c(4kl-j^2)$ to be non-zero.
Furthermore the inequality \refb{esd9} is saturated only if $k=l$
and $j^2 = 4kl+1=4k^2+1$. Since $j,k,l\in\ZZZ$, the only solution
to this is $k=l=0$, $j=\pm 1$. This contradicts our assumption
that $(k,l)\ne (0,0)$. Thus we see that the bound cannot be saturated
and we must have
\be \label{esd10}
(k+l)^2 \ge j^2\, .
\ee
Since $k,l$ are non-negative integers this gives $k+l \ge |j|$. Thus
$l'$ given in \refb{esd8} is a non-negative integer.

Next we shall examine the range of
$j'$ for $k'=l'=0$. From \refb{esd8} we see that in this case $k=0$,
$j=l$ and $j'=j=l$. Furthermore $l$ cannot vanish, since for $k=l=0$,
$j$ must be a negative integer and hence we cannot have $j=l$. This
shows that $l$ and hence $j'=l$ must be a positive integer.
Thus the
allowed range of values of $k'$, $l'$ and $j'$ takes the form
\be \label{esd10.5}
k'\ge 0, \quad l'\ge 0, \quad j'\in\ZZZ, \quad  
\hbox{$j'>0$ for $k'=l'=0$}\, .
\ee
This coincides with the allowed range of values 
of $k$, $l$ and $j$ after flipping the sign of $j$ for
$k=l=0$. As argued before, the latter operation preserves the
form of $\cps$.

In order to complete the proof of
invariance of $\cps$ under the transformation
$T$ we must now verify that all the values in the
range \refb{esd10.5} are realized, \i.e.\
given any triplet $(k',l',j')$ 
subject to the
condition \refb{esd10.5}
we can find a triplet $(k,l,j)$  satisfying
\refb{esd8} within the allowed range. 
For this we need to invert \refb{esd8} as
\be \label{esd11}
k=k', \quad l=k'+l'+j', \quad j = j'+2k'\, ,
\ee
and repeat our arguments in the opposite direction. This is a
straightforward exercise and we find that for any triplet
$k'$, $l'$ and $j'$
satisfying \refb{esd10.5}, $k$ and $l$ given in
\refb{esd11} are non-negative integers,
$j$ is an integer, and for $k=l=0$, $j$ is a positive integer
which can be turned into a negative integer using the
$j\to -j$ symmetry of the summand. This
completes the proof of invariance of $\cps$ under the 
S-duality transformation $T$.

This establishes S-duality invariance of $\cps$ given in \refb{esdual}.
Note that inclusion of the $k=0$ terms in \refb{ek6}, which
count the contribution from 1/8 BPS dyons with vanishing 
D1-brane charge, is essential for getting an S-duality invariant
partition function. Without this term the range of $k'$, $l'$ and $j'$
will not coincide with that of $k$, $l$ and $j$.

\sectiono{Wall crossing} \label{swall}

In this section we shall carry out an independent computation of
the change in the index of a 1/8 BPS dyon as we cross a wall of
marginal stability at which the 1/8 BPS dyon breaks up into a
pair of half-BPS dyons, and  compare the result with the
result predicted from \refb{ek6}.

Let us first compute the jump in the index predicted by \refb{ek6}.
According to eqs.\refb{e2}-\refb{e4},
as we cross the wall of marginal stability associated
with
the decay $(Q,P)\to (Q,0)+(0,P)$,
the contour will cross the pole at $\cv=0$. Such poles of
$\cps$ come from the $k=l=0$ term in \refb{ek6}. Thus the
change in the index as we cross this wall is given by the residue
of the integrand in \refb{ek5} at the pole at $\cv=0$. This gives
\be \label{ejump}
\Delta d(Q,P) = (-1)^{Q\cdot P+1} \, Q\cdot P\, \wh c(-1)\, 
\delta_{Q^2,0}\, 
\delta_{P^2,0}
= (-1)^{Q\cdot P+1} \, Q\cdot P\, \delta_{Q^2,0}\, 
\delta_{P^2,0}\, .
\ee
Using S-duality invariance we see that for a more general decay of
the type given in \refb{e2}-\refb{e3} we must have
$(\alpha Q+\beta P)^2=0$ and $(\delta Q - \beta P)^2=0$ for
$\Delta d(Q,P)$ to be non-zero. Combining these with the
constraints \refb{e3} we get
\ben \label{elist}
&& \alpha = {1\over2} - {Q\cdot P\over 2\sqrt{(Q\cdot P)^2
- Q^2 P^2}}, \quad \beta = {Q^2 \over 2 \sqrt{(Q\cdot P)^2
- Q^2 P^2}}, \nonumber \\
&& \gamma = -{P^2 \over 2 \sqrt{(Q\cdot P)^2
- Q^2 P^2}}, \quad \delta = {1\over2} 
+ {Q\cdot P\over 2\sqrt{(Q\cdot P)^2
- Q^2 P^2}}\, .
\een
Since $\alpha,\beta,\gamma,\delta$ must be integers, this gives
strong constraints on $Q^2$, $P^2$ and $(Q\cdot P)^2$.
In particular 
$Q^2P^2 -(Q\cdot P)^2$, known as the discriminant, must be
negative. Furthermore for a given $(Q,P)$ there is at most a
single wall of marginal stability corresponding to
$\alpha$, $\beta$, $\gamma$, $\delta$ given by \refb{elist}.
This is an enormous simplification compared to the
corresponding results in $\NN=4$ supersymmetric string
theories\cite{0702141}.

Note that \refb{ek6} has a pole at $k\crh+l\cs+j\cv=0$ for every
$(k,l,j)$ in the range of summation given in \refb{ek6}.
Comparing this to \refb{e4} we see that the poles given in
\refb{e4} arise from the $(k,l,j)=(\gamma, -\beta,\alpha-\delta)$
term in the sum in \refb{ek6}. The restrictions \refb{e3} on
$\alpha,\beta,\gamma,\delta$ then pick out those $(k,l,j)$
for which $4kl-j^2=-1$. Can the poles associated with other values
of $(k,l,j)$ cause
further jumps in the index?
To answer this question
we note that  given that the original contour, appropriate to the
weak coupling region of the first description, has $M_1,M_2>>|M_3|
>0$, it satisfies
\be \label{eks1}
M_1,M_2>0, \qquad M_1M_2 > M_3^2\, .
\ee
Since the conditions \refb{eks1} are invariant under the 
S-duality transformation
\refb{ereplace2}, all the contours corresponding to different domains
of the moduli space satisfy \refb{eks1}.
It is now easy to
see that while deforming a contour associated with one such
$(M_1,M_2,M_3)$ to a contour associated with
another  $(M_1,M_2,M_3)$ satisfying \refb{eks1},
we never hit the pole at $k\crh+l\cs+j\cv=0$ unless
$4kl-j^2<0$, -- indeed for $4kl-j^2\ge 0$,
$\Im(k\crh+l\cs+j\cv)$ is positive on the
initial and the final contours and hence can be made to remain
positive during the deformation. 
Since $\wh c(u)=0$ for $u< -1$, this gives
$4kl-j^2=-1$ as the condition for hitting the pole.
All such poles are of the form given in \refb{e3}.

For an independent
computation of the jump in the index across the wall of marginal
stability associated with the decay $(Q,P)\to (Q,0)+(0,P)$,
we use the following 
argument\cite{0005049,0010222,0101135,0206072,
0304094,0702146,0706.3193,private}. 
Let us recall first the corresponding computation in
$\NN=2$ supersymmetric
string theories where a half BPS state breaks 4 of the 8 supersymmetries
and hence has 4 fermion zero modes. Now consider a wall of
marginal stability
where a half BPS state decays into a pair of half BPS states.
On one side
of the wall the decay
products form a bound state with large separation between the
components. This bound state ceases to exist on the other side of
the wall causing a jump in the index. Thus the change
in the index can be computed by computing the index associated
with such bound states. Although naively the system has 8 fermion
zero modes -- 4 associated with each component -- four of these
zero modes take part in the interaction which form the bound state.
Thus we are left with four fermion zero modes as is expected of a single
half-BPS state of the theory. The detailed analysis of the 
quantum mechanics leading to the formation of the bound state
yields a multiplicative factor of $(-1)^{Q\cdot P+1}\, Q\cdot P$ to
the index of these bound states, thereby leading to the formula
\be \label{eform}
\Delta d(Q,P) =
(-1)^{Q\cdot P+1}\, Q\cdot P\, d_h(Q,0) \, d_h(0,P)\, ,
\ee
for the jump in the index. Here $d_h(Q,0)$ and $d_h(0,P)$
denote the index of half-BPS states carrying charges $(Q,0)$ and
$(0,P)$ respectively and represent the contribution to the bound
state index due to the internal degeneracy of each
component.

If we consider the decay of a quarter BPS dyon of $\NN=4$
supersymmetric string theory into a pair of half-BPS dyons, then the
following argument due to Denef\cite{private}
may be used to generalize the above result.
A quarter BPS state in $\NN=4$ supersymmetric string theory breaks
12 of the 16 supersymmetries and hence has
12 fermion zero modes. On the other hand the pair of 
half-BPS states each carry 8 fermion zero modes. Of the 16
fermion zero modes associated with the pair, 4 are used up in
formation of the bound state as before, leaving us with 12 fermion
zero modes required to give rise to a quarter BPS supermultiplet.
As before supersymmetric quantum mechanics produces a degeneracy
factor of $(-1)^{Q\cdot P+1}\, Q\cdot P$ and leads
to \refb{eform} for the change in the index across a
wall of marginal stability. Note that if we had replaced one or both of
the decay products by a quarter BPS state then the resulting system
would have too many fermion zero modes and hence the index
would vanish\cite{private}. This shows that
in $\NN=4$ supersymmetric string theories the change in the index
occurs only across walls where a quarter BPS dyon decays into a pair
of half BPS dyons.

Let us now apply the same line of argument to the decay of a
1/8 BPS state in $\NN=8$ supersymmetric string theories.
As pointed out before, such a dyon is accompanied by
28 fermion zero modes. On the other hand a pair of half-BPS
states in this theory carries a total of 32 fermion zero modes of
which 4 are used up for bound state formation. Thus we are again
left with the correct number of fermion zero modes and the quantum
mechanics of bound states leads to the degeneracy factor of
\refb{eform}. We now note that for the class of charge vectors 
we are considering, -- which in the second description
involve charges carried by elementary
string states and solitons without any
D-branes charges, -- the only half BPS states are those associated with
elementary string states carrying charge $(Q,0)$ with $Q^2=0$ (\i.e.\
no left- or right-moving world-sheet excitation) and their
S-duals carrying charges $(aQ, cQ)$ with $Q^2=0$, $\gcd(a,c)=1$.
Examination of the spectrum of elementary string states shows
that there is a unique half BPS supermultiplet associated with 
the charge $(Q,0)$ and hence also with the charges
$(aQ,cQ)$. Thus we have $d_h(aQ,cQ)=\delta_{Q^2,0}$, and
in particular
\be \label{el7}
d_h(Q,0)=\delta_{Q^2,0}, \qquad d_h(0,P)=\delta_{P^2,0}\, .
\ee
Substituting this into \refb{eform} we get a result in
clear agreement with \refb{ejump}.

\medskip

\noindent {\bf Acknowledgment:} We would like to thank
Nabamita Banerjee, Shamik Banerjee,
Justin David, 
Frederik Denef, Dileep Jatkar and Yogesh Srivastava
for useful discussions.


\end{document}